\documentclass[twocolumn,prl,showpacs,superscriptaddress,preprintnumbers,amsmath,amssymb]{revtex4}

\usepackage{graphicx}
\usepackage{dcolumn}
\usepackage{bm}
\usepackage[dvips]{color}

\begin{document}
\title{Strong spin-orbit coupling effects on the Fermi surface of Sr$_{2}$RuO$_{4}$ and Sr$_{2}$RhO$_{4}$}

\author{M.W. Haverkort}
\affiliation{II. Physikalisches Institut, Universit\"{a}t zu K\"{o}ln, Z\"{u}lpicher Stra\ss e 77, 50937
K\"{o}ln, Germany}
\author{I.S. Elfimov}
\affiliation{AMPEL, University of British Columbia, Vancouver, British Columbia, Canada V6T\,1Z4}
\author{L.H. Tjeng}
\affiliation{II. Physikalisches Institut, Universit\"{a}t zu K\"{o}ln, Z\"{u}lpicher Stra\ss e 77, 50937
K\"{o}ln, Germany}
\author{G.A. Sawatzky}
\affiliation{AMPEL, University of British Columbia, Vancouver, British Columbia, Canada V6T\,1Z4}
\affiliation{Department of Physics {\rm {\&}} Astronomy, University of British Columbia, Vancouver, British
Columbia, Canada V6T\,1Z1}
\author{A. Damascelli}
\affiliation{AMPEL, University of British Columbia, Vancouver, British Columbia, Canada V6T\,1Z4}
\affiliation{Department of Physics {\rm {\&}} Astronomy, University of British Columbia, Vancouver, British
Columbia, Canada V6T\,1Z1}

\date{\today}

\begin{abstract}
We present a first-principle study of spin-orbit coupling effects on the Fermi surface of Sr$_{2}$RuO$_{4}$
and Sr$_{2}$RhO$_{4}$. For nearly degenerate bands, spin-orbit coupling leads to a dramatic change of the
Fermi surface with respect to non-relativistic calculations; as evidenced by the comparison with experiments
on Sr$_{2}$RhO$_{4}$, it cannot be disregarded. For Sr$_{2}$RuO$_{4}$, the Fermi surface modifications are
more subtle but equally dramatic in the detail: spin-orbit coupling induces a strong momentum dependence,
normal to the RuO$_2$ planes, for both orbital and spin character of the low-energy electronic states. These
findings have profound implications for the understanding of unconventional superconductivity in
Sr$_{2}$RuO$_{4}$.
\end{abstract}



\maketitle

The emergence of unconventional superconductivity in Sr$_{2}$RuO$_{4}$ from a rather conventional
Fermi-liquid-like state has attracted the attention of numerous theoretical and experimental studies of this
and related materials. Although Sr$_{2}$RuO$_{4}$ is isostructural with the high-$T_{c}$ superconducting
cuprates, there seems to be little doubt that the normal state is well described by a Fermi liquid, with a
Fermi surface obtained from density-functional theory (DFT) band structure calculations \cite{LDA1,LDA2}
which agrees well with experimental determinations. The Fermi surface of Sr$_{2}$RuO$_{4}$ has been measured
with several techniques with an unprecedented accuracy: de Haas-van Alphen (dHvA \cite{dHvA}), Shubnikov-de
Haas (SdH \cite{SdH}), low and high energy angle-resolved photoemission spectroscopy (ARPES
\cite{ARPES1,ARPES2}), and Compton scattering (CS \cite{CS}) data are all available for this compound and
give a consistent outcome. The agreement of these experiments with calculations seems satisfactory but it is
not a trivial result, if one recalls the large many-body renormalization of the band-dispersion
\cite{Ingle05} and the notion that Sr$_{2}$RuO$_{4}$ can be turned into a Mott insulator by Ca doping
\cite{CaMott}.

The apparent much weaker influence of electron correlation effects in the 4$d$ ruthenates, as compared to the
3$d$ cuprates, is not unexpected. The radial extent of the $4d$ wavefunctions with one radial node is much
larger than the zero-node 3$d$ one, leading to a larger one-electron bandwidth and a smaller on-site Coulomb
repulsion. This fundamental difference between $3d$ and $4d$ transition metal compounds also explains the
dominance of the $t_{2g}$ band and low-spin-like behavior in describing the low-energy scale physics for
systems with less than 6 $d$-electrons. In other words, Hund's rule splittings are eclipsed by crystal-field
and band-structure effects in $4d$ systems. All this would suggest that Sr$_{2}$RhO$_{4}$, with one extra
$4d$-$t_{2g}$ electron, should also be well described by band theory and could provide an interesting
approach to study the behavior of the very unconventional superconducting state of Sr$_{2}$RuO$_{4}$ upon
electron doping. Although one would expect the Fermi surface to change with the addition of one extra
electron to the Fermi sea, it came as a surprise that band theory -- within an approach equivalent to that
used for Sr$_{2}$RuO$_{4}$ -- failed badly in describing the Fermi surface of Sr$_{2}$RhO$_{4}$, as
determined experimentally to a high degree of precision by both ARPES \cite{Baumberger06,Kim06} and dHvA
experiments \cite{perry}.

The subject of this paper is to understand the reason for this discrepancy. It has been suggested that the
failure of DFT in the case of Sr$_{2}$RhO$_{4}$ is a consequence of many-body interactions
\cite{Baumberger06,Kim06}. This seems however an unlikely scenario since, on general grounds, many-body
effects are expected to be of the same magnitude in  Sr$_{2}$RhO$_{4}$ and Sr$_{2}$RuO$_{4}$. Indeed, as
confirmed by experimental determinations of mass enhancement \cite {dHvA,Baumberger06} and quasiparticle
dispersion \cite {Ingle05,Baumberger06}, the renormalization parameters are closely comparable for the two
compounds and, at most, would suggest slightly larger renormalization effects for Sr$_{2}$RuO$_{4}$. We will
show that electron correlations are not the main driving force here; instead, the problem can be well
understood by taking into account the interplay between spin-orbit coupling (SOC) and the details of the band
dispersion close to the Fermi energy. In addition, the inclusion of SOC leads to a non-trivial, hitherto
unknown, momentum dependence for orbital and spin characteristics of the near-$E_F$ electronic states in
Sr$_{2}$RuO$_{4}$; their knowledge is at the very heart of the microscopic understanding of spin-triplet,
possibly orbital-dependent, superconductivity \cite{pwave}.
\begin{figure*}
\includegraphics[width=.75\linewidth]{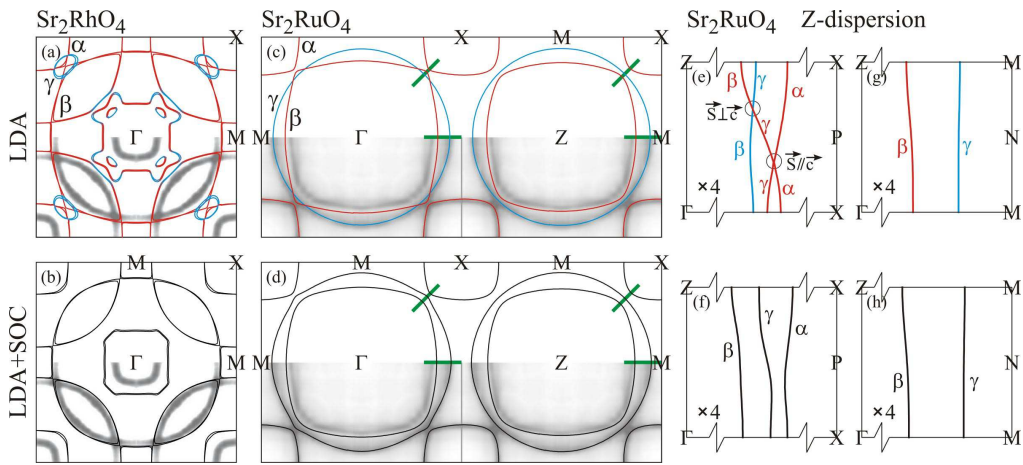}
\vspace{-.3cm}\caption{(color online). (a,b) LDA Fermi surface of Sr$_{2}$RhO$_{4}$ and (c,d)
Sr$_{2}$RuO$_{4}$; (e-h) $k_z$ electronic dispersion for the Sr$_{2}$RuO$_{4}$ cuts highlighted in (c,d) by
green solid bars. Calculations were performed without SOC (top panels: a,c,e,g) and with SOC (bottom panels:
b,d,f,h). The grey-scale ARPES data are reproduced from (a,b) Kim \textit{et al.} \cite{Kim06} and (c,d)
Damascelli \textit{et al.} \cite{ARPES1}.} \label{fig1}
\end{figure*}

Experimentally it was shown that SOC does play an important role in insulating ruthenates \cite{Mizokawa01}.
This is not surprising since the atomic relativistic SOC constant is $\zeta$=161 (191) meV for Ru$^{4+}$
(Rh$^{4+}$) \cite{Earnshaw61,SOCexplain}, which is not a small quantity. Furthermore, it is also rather
similar for the two systems leading one to believe that, if SOC is not important for the metallic ruthenates,
the same should hold for the rhodates and viceversa. As we will show, the inclusion of SOC dramatically
improves the agreement between DFT and experimental results for Sr$_{2}$RhO$_{4}$; we will also explain why
SOC has a much smaller apparent effect for Sr$_{2}$RuO$_{4}$. Care should be taken, however, since for
Sr$_{2}$RuO$_{4}$ there are specific regions in the three-dimensional Brillouin zone where SOC does have a
very large effect; this might have an important influence on the interpretation of several experimental
findings, from the details of the $k_z$ electronic dispersion \cite{dHvA,SdH} to the signatures of
spin-triplet superconducting pairing \cite{luke,kapitulnik}.

DFT calculations were done with the code WIEN2k \cite{Wien2k} and structural data reported in
Ref.\,\onlinecite{structure}. To test the generality of the results, we used several values for the radius of
the muffin-tin spheres (RMT) and two different functionals, local density approximation (LDA) and generalized
gradient approximation (GGA). The results are basically independent of RMT and extremely similar for LDA and
GGA, especially close to the Fermi energy. The plots and values shown in the paper were calculated within LDA
with RMT$_{Sr}$=1.9$a_0$, RMT$_{Ru}$=1.75$a_0$, RMT$_{O}$=1.45$a_{0}$. We used a cutoff parameter
$RK_{max}\!=\!9.0$; eigenstates with energies up to 10\,Ryd were included. Self-consistency was obtained with
$\sim$400 and $\sim$100 $k$-points in the reduced Brillouin zone for Sr$_{2}$RuO$_{4}$ and Sr$_{2}$RhO$_{4}$,
respectively. The Fermi surface was calculated on a square $k$-mesh, with points separated by
$\frac{1}{64}\pi$ within the plane of the calculated surface.

In Fig.\,\ref{fig1}a and \ref{fig1}c we show the Fermi surface of Sr$_{2}$RhO$_{4}$ and Sr$_{2}$RuO$_{4}$
calculated without SOC, which is in good agreement with previous theoretical results
\cite{LDA1,LDA2,Baumberger06,Kim06}. The Fermi surface of Sr$_{2}$RuO$_{4}$ consists of three sheets,
labelled $\alpha$, $\beta$ and $\gamma$. The $\gamma$ sheet is mainly composed of $d_{xy}$ orbitals and is
highly two dimensional; the $\alpha$ and $\beta$ sheets are formed by the one-dimensional-like bands derived
from $d_{xz}$ and $d_{yz}$ orbitals, which exhibit anticrossing behavior (i.e. mixing) along the zone
diagonal. In Fig.\,\ref{fig1}c we present calculations for both $k_z=0$ and $\pi$ (centered around $\Gamma$
and $Z$, respectively), which show some degree of $k_z$ dispersion. As $k_z$ is a difficult to determine
quantity in the experiment \cite{ARPES1}, the comparison between LDA and ARPES should be attempted for
various values of $k_z$. For $k_z=\pi$, the agreement between ARPES \cite{ARPES1}, dHvA \cite{dHvA} (not
shown), and LDA is rather good; for $k_z=0$, however, the crossing between $\gamma$ and $\beta$ sheets in LDA
is experimentally not reproduced. Note that this is really a crossing and not an anticrossing, as for $k_z=0$
the bands arising from $d_{xy}$ and $d_{xz}$/$d_{yz}$ orbitals are of different symmetry and no mixing is
allowed. Also, contrary to the LDA results of Fig.\,\ref{fig1}c, quantum oscillation experiments
\cite{dHvA,SdH} indicate very little $k_z$ modulation for the Fermi surface (smaller than 1\% of the zone
even for the $\beta$ sheet, which exhibits the largest $k_z$ effects).

In principle the same discussion holds for Sr$_{2}$RhO$_{4}$; there are however a few complications due to
structural distortions. The RhO$_{6}$ octahedra in Sr$_{2}$RhO$_{4}$ are rotated around the $c$-axis, which
results in a $\sqrt{2}\times\sqrt{2}$ doubling of the unit cell in the $a$-$b$ plane, and also a doubling
along the $c$-axis because of the alternation in the RhO$_{6}$ rotation direction. This leads to a doubling
of all bands and Fermi sheets, as evidenced in Fig.\,\ref{fig1}a by the back-folding of the Fermi surface
sheets with respect to the line connecting two closest $M$ points, e.g. ($\pi$,0,0) and (0,$\pi$,0). In
addition, this distortion reduces the electronic bandwidth and allows mixing between $d_{xy}$ and
$d_{x^2-y^2}$ orbitals \cite{Kim06}, which causes the $\gamma$ sheet to almost disappear; small remaining
$\gamma$ pockets can be seen around the $\alpha$ and $\beta$ sheet crossings. The agreement between LDA and
experiment for the rhodate is, as shown in Fig.\,\ref{fig1}a and also noted in the literature
\cite{Baumberger06,Kim06}, not very good (especially if compared with the beautiful agreement found for
Sr$_{2}$RuO$_{4}$).

At present, there is no explanation of why LDA is able to describe Sr$_{2}$RuO$_{4}$ but fails for Sr$_{2}$RhO$_{4}$, two materials that are
structurally and electronically quite similar. We will show here, starting from the more striking case of Sr$_{2}$RhO$_{4}$, that a resolution
of this discrepancy can be obtained with the inclusion of SOC in LDA. As one can see in Fig.\,\ref{fig1}b, the agreement between LDA+SOC and
experiment is very satisfactory for Sr$_{2}$RhO$_{4}$: the $\gamma$ sheet is now completely absent, and $\alpha$ and $\beta$ sheets are
significantly smaller (the $\alpha$ pocket is still somewhat larger than in the experiment, which could possibly be solved by fine-tuning the
RuO$_{6}$ distortion at the surface \cite{Kim06} but is beyond the point here). As for Sr$_{2}$RuO$_{4}$ (Fig.\,\ref{fig1}d), SOC effects are
much smaller, especially for $k_z=\pi$; some improvement can be observed only for $k_{z}=0$, in which case the $\gamma$ and $\beta$ sheet
crossing becomes an anticrossing because the bands no longer have different symmetry. This point is important, as we will stress below, but
first we discuss the reason why the effect of SOC for Sr$_{2}$RhO$_{4}$ seems so much stronger than for Sr$_{2}$RuO$_{4}$.

\begin{figure}
\includegraphics[width=.80\linewidth]{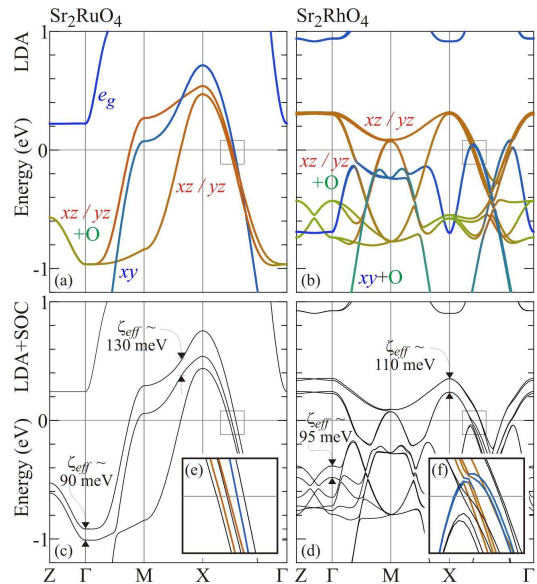}
\vspace{-.3cm}\caption{(color online). (a,c,e) Sr$_{2}$RuO$_{4}$ and (b,d,f) Sr$_{2}$RhO$_{4}$ LDA dispersion
calculated without SOC (a,b) and with SOC (c-f). (e,f) Enlarged view of the marked $E_F$-regions in (c,d).}
\label{fig2}
\end{figure}

Fig.\,\ref{fig2} presents the band structure of Sr$_{2}$RuO$_{4}$ and Sr$_{2}$RhO$_{4}$ from LDA (top) and
LDA+SOC (bottom). It is clear that if the $t_{2g}$ bands are already well separated in energy, e.g. at the M
point for Sr$_{2}$RuO$_{4}$ in Fig.\,\ref{fig2}a, SOC only mildly shifts the LDA eigenstates. A whole new
energy splitting is instead induced by SOC at those momenta characterized, in LDA, by degenerate $t_{2g}$
bands. For instance in Fig.\,\ref{fig2}c, the degenerate $d_{xz}^{(\uparrow,\downarrow)}$ and
$d_{yz}^{(\uparrow,\downarrow)}$ LDA bands of Sr$_{2}$RuO$_{4}$ along $\Gamma$-$Z$ are recombined into two
doubly-degenerate complex bands [$d_{1}^{\downarrow}$,$d_{-1}^{\uparrow}$ and
$d_{-1}^{\downarrow}$,$d_{1}^{\uparrow}$ with $d_{\pm1}=\mp\frac{1}{\sqrt{2}} (d_{xz}\pm\imath d_{yz})$],
split by an energy equal to the SOC constant $\zeta\!=\!93$\,meV. Note that the latter is derived from the
splitting at the $\Gamma$ point, and the strong reduction with respect to the 161\,meV atomic value of Ru is
mainly due to the dilution effect resulting from the covalent mixing with the O-$2p$ bands (accordingly, one
also finds a SOC splitting for the O-$2p$ derived bands around -3.8 eV due to the mixed-in Ru character; and
the SOC splittings are larger at $X$ and $M$ points, where the covalent mixing is smaller). One might wonder,
in this regard, whether the larger SOC effects in Sr$_{2}$RhO$_{4}$ might arise because of the $\sim\!20$\%
larger atomic SOC for Rh (191\,meV) as compared to Ru, originating from the somewhat less extended $4d$
orbitals. However, the on-site energy difference for Rh-$4d$ and O-$2p$ orbitals is smaller than for Ru-$4d$
and O-$2p$, making Sr$_{2}$RhO$_{4}$ more covalent than Sr$_{2}$RuO$_{4}$. As a result, the {\it effective}
SOC for the $t_{2g}$ bands in the two compounds is about equal (Fig.\,\ref{fig2}c,d); the magnitude of the
atomic SOC constants is thus not the reason for the different behavior.

More clues on the magnitude of SOC effects for the two compounds come from the inspection of the
Sr$_{2}$RhO$_{4}$ band dispersion in Fig.\,\ref{fig2}b,d. The structural distortion responsible for band
folding and narrowing in Sr$_{2}$RhO$_{4}$ gives rise to several nearly-degenerate states close to the Fermi
level (e.g., the $d_{xy}$ band maximum at $X$ in Fig.\,\ref{fig2}a is back-folded to the now equivalent
$\Gamma$ point, leading to the mixing with the $e_g$ band and the appearance of two $d_{xy}$-derived maxima
just above $E_F$, between $\Gamma$ and $X$ in Fig.\,\ref{fig2}b).
Because of the presence of weakly-dispersive bands close to $E_F$, the SOC-induced splitting causes substantial modifications of the Fermi
surface in Sr$_{2}$RhO$_{4}$ (yet in the full respect of Luttinger's theorem), such as the complete disappearance of the $\gamma$ sheet
(Fig.\,\ref{fig2}f). On the contrary in the case of Sr$_{2}$RuO$_{4}$ in the very same $k$-space region (Fig.\,\ref{fig2}e), the energy shift
from SOC does not translate in large changes in $k$ because of the steep band dispersion. It is therefore the interplay between SOC and band
filling (i.e., the presence, or lack thereof, of weakly-dispersive degenerate states close to the chemical potential), which determines the more
dramatic Fermi surface changes for Sr$_{2}$RhO$_{4}$ in LDA+SOC. Larger effects should be expected for Sr$_{2}$RuO$_{4}$ upon increasing the
electron filling; this has indeed been observed in Sr$_{2-y}$La$_y$RuO$_4$ \cite{kyle}, in which the $\gamma$-sheet topological transition
detected for $y\!\gtrsim\!0.20$ is well accounted for by LDA+SOC (not shown).

In the following, we will focus on some aspects of the LDA+SOC results which, beyond the detailed analysis of
the Fermi surface, promise to have profound implications for the physical properties of Sr$_{2}$RuO$_{4}$. We
have already noted, in relation to Fig.\,\ref{fig1}c, that for $k_z=0$ the LDA calculations display the
crossing of $\gamma$ and $\beta$ sheets at eight momenta symmetrically located with respect to the
$\Gamma$-$X$ lines. As SOC is not a small perturbation at highly-degenerate $k$-points in LDA, to understand
the impact of SOC on the low-energy physics we have to follow this degeneracy as a function of $k_z$. In
Fig.\,\ref{fig1}e,g we plot the $k_z$ modulation of the LDA Fermi sheets in the $\Gamma$-$X$-$Z$ and
$\Gamma$-$M$-$Z$ planes. The $k$-space loci of $\gamma$ and $\beta$ sheet degeneracy in LDA defines an
elliptical contour centered around the $\Gamma$-$X$ line, crossing the $k_z=0$ plane at
{($k_x$,$k_y$)}=(0.28$\pi$,0.34$\pi$) and (0.34$\pi$,0.28$\pi$) as indicated in Fig.\,\ref{fig1}c, and
crossing the $\Gamma$-$X$-$Z$ plane at $k_z=0.68\pi$. This is a degeneracy between the $d_{xy}$ orbital and a
linear combination of $d_{xz}$ and $d_{yz}$ orbitals; there is also a further degeneracy at
($k_x$,$k_y$,$k_z$)=(0.34$\pi$,0.34$\pi$,$0.33\pi$) between the $d_{xz}$ and $d_{yz}$ orbital
(Fig.\,\ref{fig1}e). In addition to lifting the degeneracy of the LDA Fermi surfaces (Fig.\,\ref{fig1}f), the
inclusion of SOC reduces to more realistic values the $k_z$-modulation of each individual Fermi surface
sheet, which in LDA is much larger than observed in quantum oscillation experiments \cite{dHvA,SdH}. Note, in
this regard, that the rather localized bulging of the Fermi surfaces in LDA+SOC, such as e.g. along the zone
diagonal for $k_z\!=\!0$ for the $\gamma$ sheet, might require the inclusion of higher order cylindrical
harmonics than usually considered in the quantum oscillation analysis \cite{dHvA,SdH}.

Perhaps more important, SOC is responsible for the three-dimensional $k$-dependence of the wavefunction spin and orbital character, along the
Fermi surface. For instance, the character of the $\gamma$ sheet is no longer pure $d_{xy}$; instead, for $k_x=k_y$ and $|k_z|<0.68\pi$, it is
mainly a linear combination of $d_{xz}$ and $d_{yz}$. This not only influences the hopping parameters; it also has a large effect on the spin
anisotropy. At (0.34$\pi$,0.34$\pi$,$0.33\pi$), SOC recombines the degenerate $d_{xz}^{(\uparrow,\downarrow)}$ and
$d_{yz}^{(\uparrow,\downarrow)}$ orbitals into the two doubly-degenerate $d_{1}^{\downarrow}$,$d_{-1}^{\uparrow}$ and
$d_{-1}^{\downarrow}$,$d_{1}^{\uparrow}$ complex orbitals. These new eigenstates have a spin-moment in the $z$ direction and a zero expectation
value for  $x$ and $y$ spin components. This follows from $s_x=\frac{1}{2}(s^++s^-)$ and $\langle
d_{1}^{\downarrow}|s^-|d_{-1}^{\uparrow}\rangle=0$, as the orbitals are orthogonal to each other, and analogous relations for $s^+$ and $s_y$.
Thus, at (0.34$\pi$,0.34$\pi$,$0.33\pi$) the spins on the $\gamma$ and $\alpha$ sheets point along the $c$-axis, as indicated in
Fig.\,\ref{fig1}e. Similar arguments can be made if the $d_{xy}$ and $d_{xz}$ or $d_{yz}$ orbitals are degenerate. By cyclic permutation of
coordinates, it can be shown that this leads to a spin direction in the $a$-$b$ plane (Fig.\,\ref{fig1}e), along the whole elliptical
LDA-degeneracy contour between $\gamma$ and $\beta$ sheets discussed above. Note that this $k$-dependent non-collinear spin anisotropy cannot
easily be seen in the bulk magnetic susceptibility, which measures an average over all $k$-points. It has however severe implications for all
$k$-dependent properties of Sr$_{2}$RuO$_{4}$.

Our study shows that, when describing experiments on Sr$_{2}$RuO$_{4}$ and Sr$_{2}$RhO$_{4}$, it is crucial to start from accurate ab-initio
band-structure calculations including SOC. One should be careful about a phenomenological modelling of the electronic structure. For instance, a
simple tight-binding fit to the experimentally determined Fermi surface of Sr$_{2}$RuO$_{4}$ leads to non-degenerate bands at the Fermi energy;
adding SOC, starting from such a scenario, would only introduce a small perturbation. SOC is however not a small perturbation for metals if
(nearly) degenerate bands cross or are close to the Fermi energy, which happens for Sr$_{2}$RhO$_{4}$ in extended portions of the Brillouin zone
and for Sr$_{2}$RuO$_{4}$ at some momenta. This has important implications -- macroscopic in the case of Sr$_{2}$RhO$_{4}$ -- for fundamental
properties such as the detailed shape and orbital character of the Fermi surface sheets, as well as the amount of $k_z$ dispersion. For example,
SOC is responsible for a significant reduction of the effective three-dimensionality in Sr$_{2}$RuO$_{4}$, which is consistent with the dHvA
results \cite{dHvA}. Also, even in the overall less dramatic case of Sr$_{2}$RuO$_{4}$, SOC leads to a spin anisotropy and non-collinear
behavior in several $k$-space regions, with the spin precession direction varying with $k$ according to the orbital character of the LDA+SOC
eigenstates. This $k$-dependent orientation of the expectation value of the spin calculated on the LDA+SOC eigenstates, which reminds us of
Rashba coupling, could be a very important ingredient in describing the unconventional superconducting state of Sr$_{2}$RuO$_{4}$ and should be
investigated further. Since all Fermi surface degeneracies are lifted in LDA+SOC and the $k$-dependent energy splittings are much larger then
the superconducting pairing energy, the SOC-induced spin anisotropy in concert with orbital mixing should directly affect the orbital and spin
angular momentum of the Cooper pairs. In fact, singlet and triplet states could be strongly mixed, blurring the distinction between spin-singlet
and spin-triplet pairing \cite{pwave}, and with it the simple interpretation of time-reversal symmetry-breaking experiments
\cite{luke,kapitulnik}.

This work was supported by the Alfred P. Sloan Foundation (A.D.), CRC and CIFAR Quantum Materials Programs
(A.D., G.A.S.), NSERC, CFI, BCSI, the Deutsche Forschungsgemeinschaft (DFG) through SFB 608, and NSF under
Grant No. PHY05-51164.


\begin{thebibliography}{99}
\bibitem{SOCexplain}The SOC Hamiltonian is $\sum_i \zeta l_i \cdot s_i$, where $l_i$ ($s_i$) is the orbital (spin) angular-momentum
operator for the $i$-th electron, and the sum runs over all $i$ electrons.
\end{thebibliography}
\end{document}